\documentclass[conference]{IEEEtran}
\IEEEoverridecommandlockouts

\usepackage{cite}
\usepackage{amsmath,amssymb,amsfonts}
\usepackage{algorithmic}
\usepackage{graphicx}
\usepackage{textcomp}
\usepackage{xcolor}
\usepackage{comment}
\usepackage{xurl}
\usepackage{csquotes}

\graphicspath{{figures/}{pictures/}{images/}{./}} 

\usepackage{times}                     

\usepackage{tabu}                      
\usepackage{booktabs}                  
\usepackage{lipsum}                    
\usepackage{mwe}        
\usepackage{comment}

\usepackage{mathptmx}                  

\usepackage{float}
\usepackage{csquotes}
\usepackage[flushleft]{threeparttable}
\usepackage{tikz}

\def\BibTeX{{\rm B\kern-.05em{\sc i\kern-.025em b}\kern-.08em
    T\kern-.1667em\lower.7ex\hbox{E}\kern-.125emX}}
\begin{document}

\title{Exploring Augmented Table Setup and Lighting Customization in a Simulated Restaurant to Improve the User Experience
}

\author{
 \IEEEauthorblockN{Jana Motowilowa$^1$, Maurizio Vergari$^1$, Tanja Kojić$^1$, Maximilian Warsinke$^1$, Sebastian Möller$^{1,2}$, Jan-Niklas Voigt-Antons$^3$}
 \IEEEauthorblockA{$^1$TU Berlin, Germany, $^2$DFKI, Germany, $^3$Hochschule Hamm-Lippstadt, Germany}
}

\maketitle

\begin{abstract}
This study explored a concept for using \textit{Augmented Reality} (AR) glasses to customize augmented table setup and lighting in a restaurant. The aim was to provide insights into AR usage in restaurants and contribute to existing research by introducing an extendable and versatile concept for scholars and restaurateurs. A controlled laboratory study, using a within-subjects design, was conducted to investigate the effects of a customizable augmented table setup and lighting on \textit{user experience} (UX), \textit{perceived waiting time}, \textit{psychological ownership} and \textit{social acceptability}. A simulated restaurant environment was created using 360$^{\circ}$ image in \textit{Virtual Reality} (VR). The study implemented default and customizable \textit{table setup} and \textit{lighting}. Results from a paired samples t-test showed a statistically significant effect of table setup and lighting on \textit{pragmatic quality of UX}, \textit{hedonic quality of UX}, \textit{overall UX}, \textit{Valence}, \textit{Dominance}, \textit{psychological ownership} and \textit{affect}. Furthermore, table setup had a significant effect on \textit{Arousal} and \textit{perceived waiting time}. Moreover, table setup significantly affected \textit{AR Interaction-}, \textit{Isolation-}, and \textit{Safety acceptability}, while lighting only affected \textit{AR Interaction acceptability}. Findings suggest that these investigated variables are worth considering for AR applications in a restaurant, especially when offering customizable augmented table setup and lighting.
\end{abstract}

\newcommand\copyrighttext{%
  \footnotesize \textcopyright\ 2024 IEEE. Personal use of this material is permitted. Permission from IEEE must be obtained for all other uses, in any current or future media, including reprinting/republishing this material for advertising or promotional purposes, creating new collective works, for resale or redistribution to servers or lists, or reuse of any copyrighted component of this work in other works. DOI and link to the original publication will be added as soon as they are available}

\newcommand\copyrightnotice{%
\begin{tikzpicture}[remember picture,overlay,shift={(current page.south)}]
  \node[anchor=south,yshift=10pt] at (0,0) {\fbox{\parbox{\dimexpr\textwidth-\fboxsep-\fboxrule\relax}{\copyrighttext}}};
\end{tikzpicture}%
}

\copyrightnotice


\section{Introduction and Related Work}

Augmented Reality (AR) and Virtual Reality (VR) technologies are rapidly growing and are expected to reach 3.728 million people worldwide by 2029 \cite{statista}. While immersive technologies were adopted across various fields such as e-commerce and gaming \cite{statista}, their use in restaurants is still uncommon. Present use cases include immersive experiences \cite{61batat2021,68kanak2018,69caboni2021, 6RWang2021}, interactive menus, and 3D representation of dishes \cite{16fritz2023, onirix,2Rpetit2022} to enhance the dining experience. However, there is an additional application of AR in the restaurant context: allowing customers to personalize aspects of their environment.

This study delves into the potential of AR to enable positive emotional states and hedonic benefits through a customizable augmented table setup and lighting in a simulated restaurant environment in VR. Prior research has successfully utilized VR to simulate AR across various contexts, showing minimal differences between real-world and simulated conditions \cite{92lacoche2022, 94alce2015, 93pfeiffer-lessmann2018}. Also, studies showed that the physical environment in restaurants and its interior design are important factors for customer experience \cite{98walter2012,99chen2019}, customer emotion \cite{99chen2019, liu2009effects,36hwang2012,4Rmotoki2021}, and customer behavior \cite{96pecotic2014, 97spence2012, liu2009effects, 36hwang2012, 5Rspence2020,4Rmotoki2021}. The relevance of the table setup and lighting in a restaurant is given by them being an element of the restaurant's interior and environment. In this study, \textit{table setup} refers to the virtual objects on the dining table, while \textit{lighting} refers to the illumination of the simulated restaurant environment. 

Examining the social acceptability of new emerging technologies such as AR glasses in public use cases is essential to determine their future success and adoption rate \cite{vergari2021, 129alallah,88daiin2019, 89denning2014, 87koelle2015}. \textit{Social acceptability} can be defined as the phenomenon of judging a technology introduced in the future from the comfort or discomfort of both the user's and the observer's perspective. It can be evaluated through various dimensions including \textit{Public VR}, \textit{User}, \textit{Public Communication}, \textit{VR Interaction}, \textit{Isolation}, \textit{Privacy}, and \textit{Safety} \cite{126eghbali2018,vergari2021}.

This work provides insights into the social acceptability of AR glasses in a restaurant context and investigates psychological ownership toward augmented table setup and lighting. \textit{Psychological ownership} can be defined as 
\enquote{the state of the mind in which individuals feel as though the object or target of ownership is theirs} \cite{44poretski2021}. It is related to relevant variables for restaurants such as \textit{customer-company identification}, \textit{word-of-mouth}, and \textit{willingness to pay more} \cite{130asatryan2008}. Relevant research on psychological ownership in AR can be found in \cite{44poretski2021, 33carrozzi2019}.

Waiting is also an intrinsic part of the restaurant experience. Investigating the perceived waiting time is relevant, given its influence on customers' evaluation of services \cite{114worlitz2020}. Additionally, studies have demonstrated that the environment in waiting situations plays a crucial role in customers' perceptions and behavior \cite{75bilgili2018, 36hwang2012}. If customizing elements of the restaurant's interior contributes to creating a favorable environment, it could prevent or mitigate a negative waiting experience \cite{36hwang2012}, which, in turn, would be reflected in the perceived waiting time. In this study, \textit{perceived waiting time} refers to the subjective time spent in the simulated restaurant environment waiting for the order, covering the in-process phase, from ordering an appetizer until its delivery \cite{114worlitz2020, 125bielen2007, 133pruyn1998, 115hwang2006}.

\begin{figure}[tb]
    \centering
    \includegraphics[width=\linewidth]{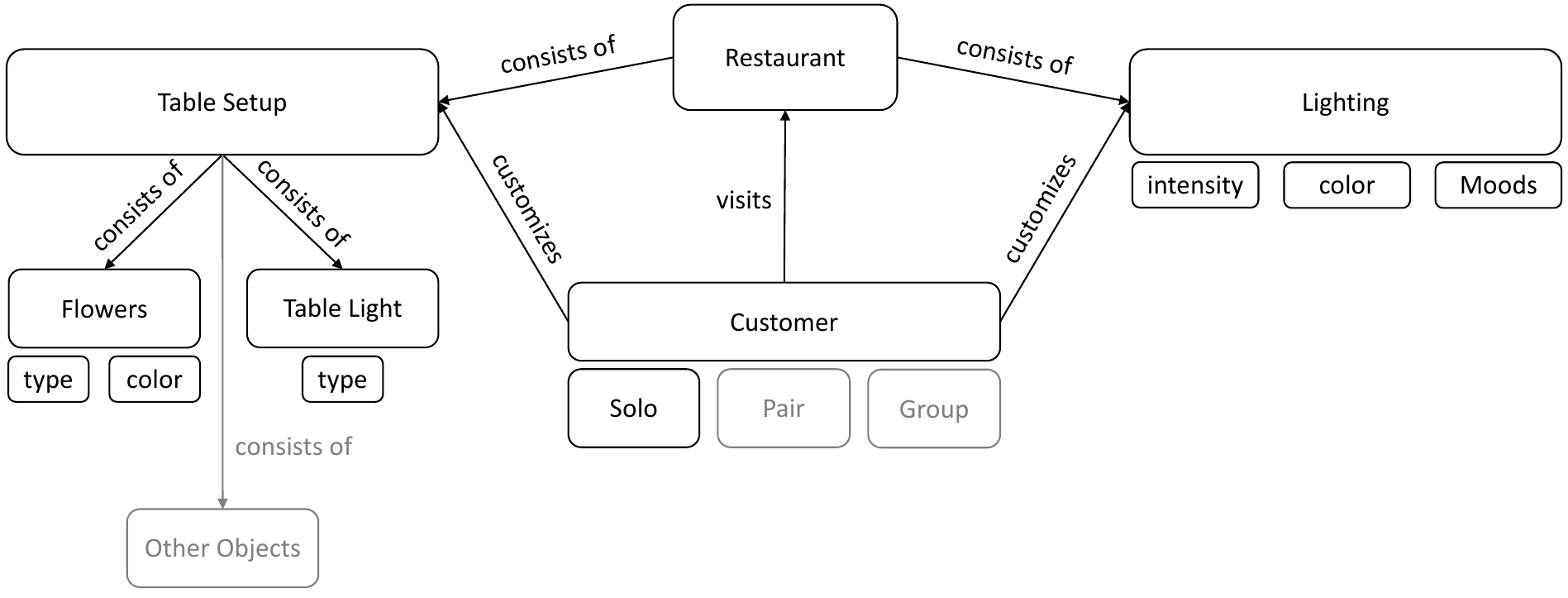}
    \caption{Concept Model}
    \label{fig:concept}
\end{figure}

The concept of this work centers around customers using AR glasses inside a restaurant to customize virtual objects on their dining table and the lighting. In the concept, the restaurant is a physical environment consisting of interior elements like the table setup and lighting. The table setup consists of decorative elements like flowers and a table light. Customers can customize table setup elements in terms of type (and color) and the lighting in terms of intensity and color. Central to this concept are the customers who can benefit from the customization, whether dining alone, in pairs, or groups. This study investigates the solo experience, shown in black in Fig. \ref{fig:concept}. A shared experience and an exemplary extension of the table setup are implied in gray. This study draws from Obrist's approach \cite{3Robrist2021}, which suggests that multisensory experiences are impressions formed by events whose sensory characteristics were crafted intentionally.

\section{Methods}
\label{sec:methods}

The research questions for this work were the following:

\begin{itemize}
    \item RQ1: What are the effects of a customizable augmented table setup on user experience, perceived waiting time, psychological ownership, and social acceptability in a simulated restaurant environment?
    \item RQ2: What are the effects of customizable augmented lighting on user experience, perceived waiting time, psychological ownership, and social acceptability in a simulated restaurant environment?
\end{itemize}

This study used a within-subject design with two pairwise comparisons. 
The two pairwise comparisons forming a simulated restaurant environment were:
\begin{itemize}
    \item \textit{Table setup} (A): default (1), customizable (2)
    \item \textit{Lighting} (B): default (1), customizable (2)
\end{itemize}

The combinations of those variables led to four conditions to be tested: A1, A2, B1, and B2. These conditions could address the research questions.
Considering the table setup pairwise comparison (A1 vs. A2), \textit{default} corresponded to wearing AR glasses and looking at the table setup including three red roses and a candle without customization options. In contrast, \textit{customizable} allowed customization of the table setup.
When it comes to the lighting pairwise comparison (B1 vs. B2), \textit{default} corresponded to wearing AR glasses and looking at an empty table and a default lighting without customization options, while \textit{customizable} provided the possibility to customize the lighting.

\subsection{Participants}
\label{sec:methods-participants}

Nineteen participants took part in the study ($8$ females, $11$ males). The average age was 26.63 years (SD = 5.90 years, min. 22 years, max. 46 years). Participants were recruited through word-of-mouth. Comprising the majority, 16 participants indicated having used a VR headset before. The average Affinity for Technology Interaction score \cite{ati} was 4.35 (SD = 0.88). All participants provided written informed consent before participating in the experiment. The user study was conducted in a controlled laboratory room of TU Berlin. The local ethics committee approved the experiment.

\subsection{Apparatus}
\label{methods-materials-apparatus}

The VR application was implemented in Unity 3D and ran on a Meta Quest 1. The hand-tracking feature was used as the input modality. The Meta Quest App streamed the user's view during the experiment. The questionnaires were built and presented on a laptop through Google Forms. The consent form and the introduction sheet were presented on paper. The introduction sheet detailed the experiment's aim, procedure, and modalities. It described the scenario of the user study being a restaurant offering an AR glasses experience. 
\subsection{Design}
\subsubsection{Table Setup}

\begin{figure}[tb]
    \centering
    \includegraphics[height = 4.10cm, width=\linewidth]{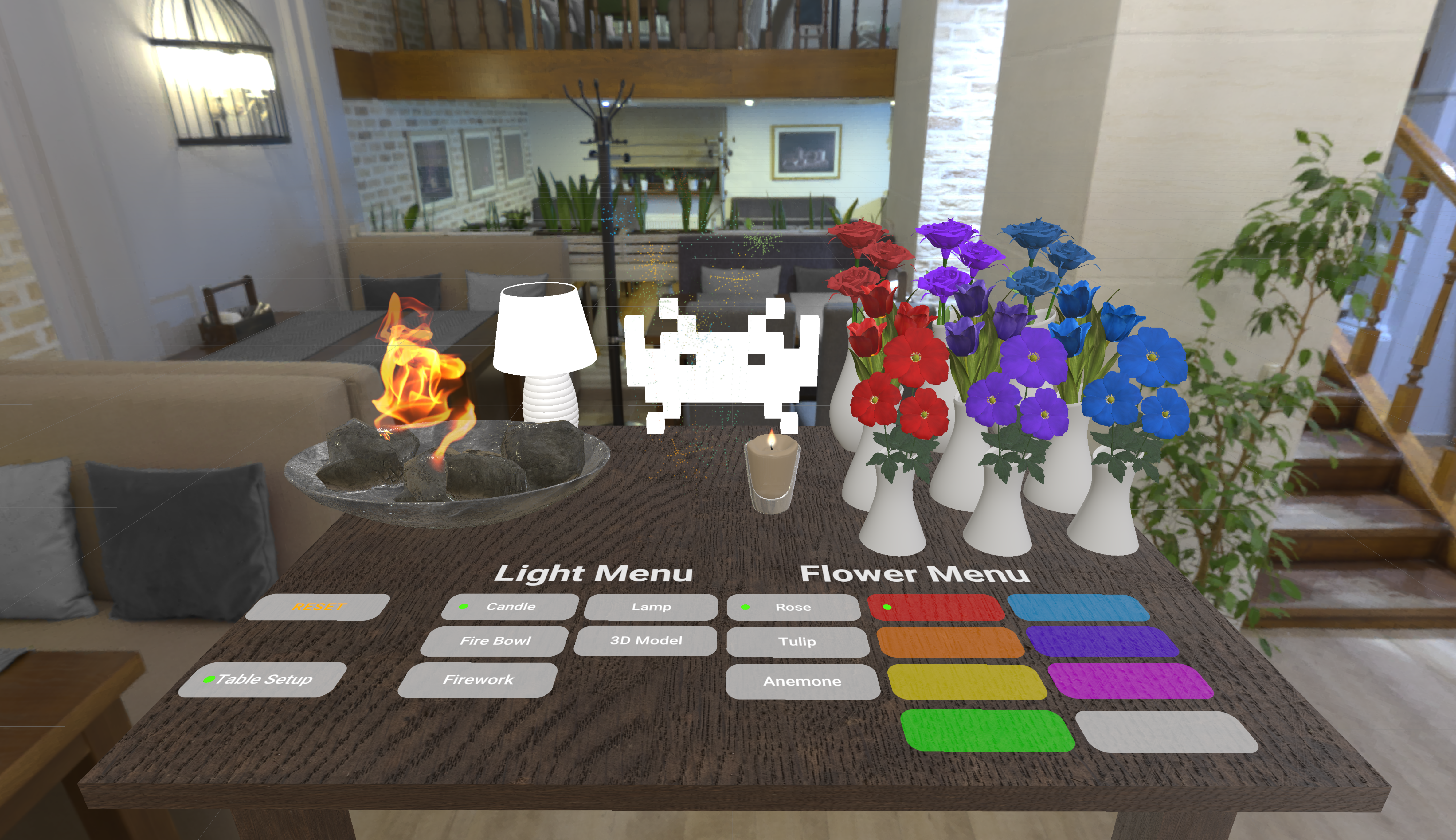}
    \caption{Table setup elements examples and UI for customizing the flower type and color, and light type in table setup customizable condition in the simulated restaurant environment.}
    \label{fig:design-table setup}
\end{figure}

In this experiment, flowers and table lights formed the table setup. Flowers were a composition of three equally colored identical flowers arranged in a vase placed on the top right of the table, and table lights were various illuminating objects placed on the top middle of the table.
The categorical attributes type and color defined a flower with the following values: \textit{Flower type}: rose, tulip, anemone; \textit{Flower color}: red, orange, yellow, green, blue, purple, pink, white.
A table light was defined by the categorical attribute type with the following values: \textit{Table light type}: candle, fire bowl, firework, lamp, 3D model. In the default condition, three red roses and a candle were presented. In the customizable condition, the user could select the flower's type and color and one of the five types of table light. Fig. \ref{fig:design-table setup} depicts all flower types in example colors and all table light types available in the experience. Furthermore, an open UI is shown, consisting of buttons placed on the dining table for a tactile response from the table the participant was sitting at when pushing through hand-tracking. A green sphere indicates the selected values. The displayed table setup arrangement in Fig. \ref{fig:design-table setup} was not in the user testing.

\subsubsection{Lighting}
\begin{figure}[tb]
    \centering
    \includegraphics[height = 3.87cm , width=\linewidth]{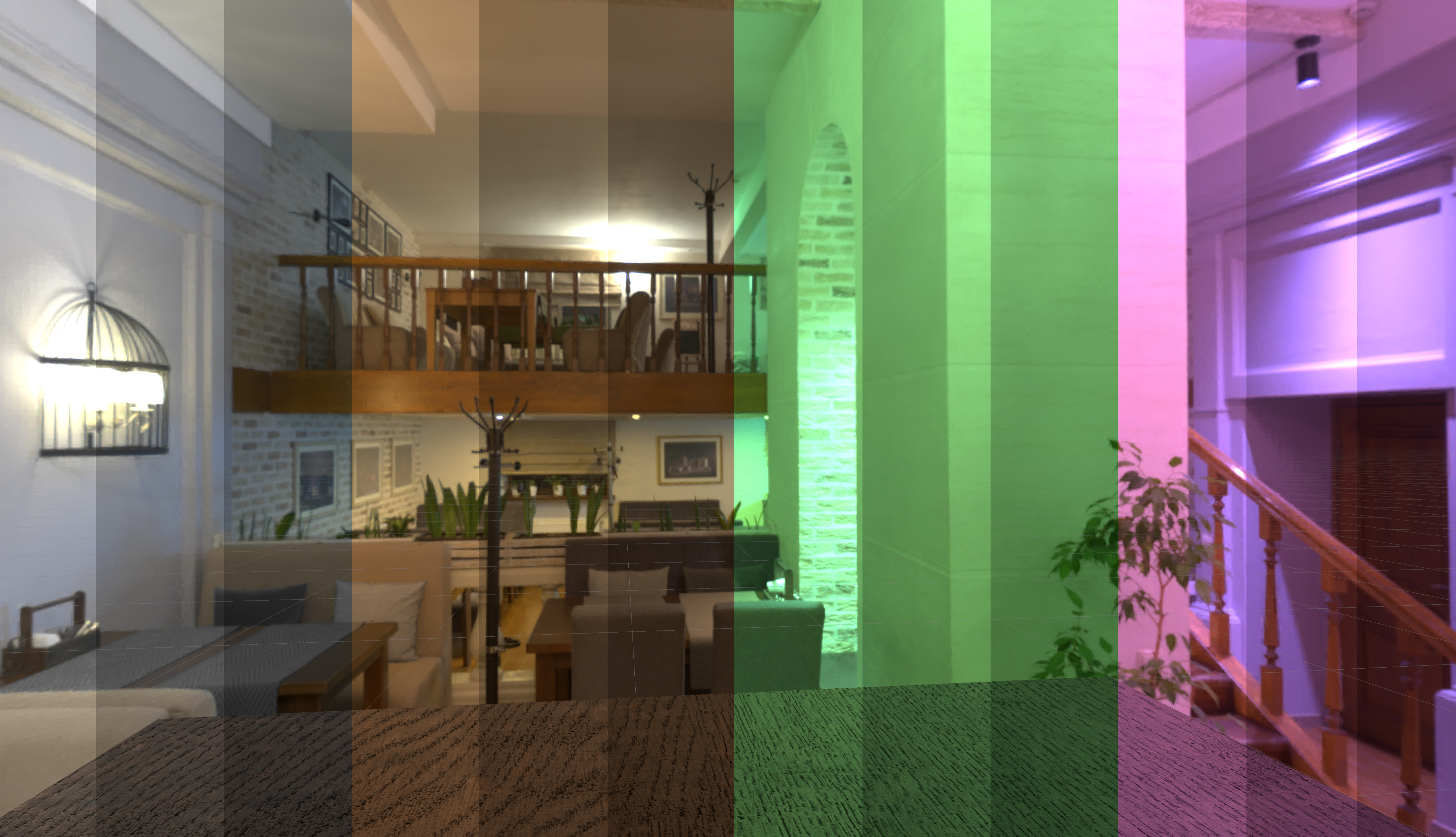}
    \caption{Collage of all lighting intensities in example colors in the lighting customizable condition in the simulated restaurant environment.}
    \label{fig:design-lighting}
\end{figure}

In this experiment, lighting was defined by the categorical attributes intensity and color, with the following values for each attribute: \textit{Lighting intensity}: bright, medium, dark; \textit{Lighting color}: red, orange, yellow, green, blue, purple, pink, white. In the default condition, the lighting was medium white. In the customizable condition, the user could select a combination of lighting intensity and color. Fig. \ref{fig:design-lighting} illustrates the lighting colors white, orange, green, and pink, each in a gradient of lighting intensity from bright, medium, and dark. 

\subsection{Procedure}

On average, the duration per participant was 66 minutes (min. 44, max. 109).
The moderator welcomed the participant and asked them to read and sign the consent form. After reading the introduction sheet, the participant completed a pre-questionnaire (demographics and ATI \cite{ati}). The participant was then instructed on how to use hand-tracking and adjust the VR headset. The participant could now see the training scene, showing buttons on a table. The participant was advised to look at their hands and push the buttons using them. After the training was completed, the first condition was started. The moderator said: \enquote{\textit{Now, please imagine the scenario you read previously: You are in a restaurant waiting for your appetizer to arrive while wearing the AR glasses. You will encounter either a default or a customizable situation}}. The moderator observed the stream. When everything worked, the moderator said: \enquote{\textit{Now, please wait for the appetizer}}. The moderator started a timer and observed the participant. In the customizable conditions, after 3 minutes, the participant was asked to set up the table or lighting as they liked best and to tell the moderator when they were done. The moderator repeated the selection and noted it in a table. After 4 minutes, the moderator said: \enquote{\textit{Your appetizer has arrived. Please take the headset off}}. The participant filled out the post-task questionnaire (PWT, SAM \cite{110bradley1994}, UEQ-S \cite{120schrepp2022}, adapted POQ \cite{33carrozzi2019}, adapted SAQ \cite{vergari2021}) while the moderator set up the next condition. These steps were repeated four times until all conditions were played and post-task questionnaires were answered. Then, the participant filled out the final questionnaire (multiple choice and open-ended questions).

\section{Results}
\label{sec: results}

\subsection{Table Setup}
\label{sec:results-table-setup}
A paired samples t-test was performed to evaluate whether there was a difference between the table setup default and customizable condition. It was found that table setup had a statistically significant effect on the dependent variables perceived waiting time (PWT), user experience (SAM and UEQ-S), psychological ownership (POQ), and social acceptability (SAQ) dimensions. An overview of effects is given in Table \ref{tab:analysis-t-test-table-setup}. The average PWT was significantly lower in the table setup customizable condition (M = 4.53, SD = 1.35) than in the default condition (M = 5.47, SD = 1.54). The table setup was significantly more pleasant (SAM Valence) in the customizable condition (M = 3.95, SD = 0.97) than in the default condition (M = 3.11, SD = 1.05). The table setup was significantly more intense in the customizable condition (M = 2.74, SD = 0.93) than in the default condition (M = 2.05, SD = 0.91) (SAM Arousal). Moreover, the average SAM Dominance rating indicated significantly higher user control in the table setup customizable condition (M = 3.58, SD = 0.77) than in the default condition (M = 3.11, SD = 0.99). The average UEQS Pragmatic value was significantly better in the table setup customizable condition (M = 1.99, SD = 0.57) than in the default condition (M = 0.68, SD = 1.19). Furthermore, the average UEQS Hedonic value was significantly better in the table setup customizable condition  (M = 1.04, SD = 1.29) than in the default condition (M = $-$0.59, SD = 1.61). Moreover, the average UEQS Overall value was significantly better in the table setup customizable condition (M = 1.51, SD = 0.84) than in the default condition (M = 0.05, SD = 1.33). Participants' feelings of ownership toward the table setup (POQ Psychological Ownership) were significantly higher in the table setup customizable condition (M = 4.51, SD = 1.53) than in the default condition (M = 2.65, SD = 1.29). Furthermore, participants' feelings toward the table setup (POQ Affect) were significantly better in the table setup customizable condition (M = 5.65, SD = 1.15) than in the default condition (M = 4.77, SD = 1.39). Participants reported significantly higher acceptability of interactions (SAQ Interaction) in the table setup customizable condition (M = 4.55, SD = 1.53) than in the default condition (M = 4.18, SD = 1.49). Furthermore, results showed that AR isolation (SAQ Isolation) was significantly more acceptable in the table setup customizable condition (M = 4.47, SD = 1.17) than in the default condition (M = 3.61, SD = 1.25). Participants felt significantly less safe (SAQ Safety) in the table setup customizable condition (M = 4.74, SD = 1.66) than in the default condition (M = 5.11, SD = 1.88). 

\begin{table}[tb]
    \centering
    \caption{Paired Samples T-Test Statistically Significant Results (\textit{p}$<$0.05) for the Independent Variable Table Setup ($N = 19$)}
        \begin{threeparttable}[t]
        \begin{tabular}{lrrrrrr}
        \toprule
        \multicolumn{1}{c}{Dependent Variable} & \multicolumn{1}{c}{$t$} & \multicolumn{1}{c}{$df$}& \multicolumn{1}{c}{$p$}& \multicolumn{1}{c}{$d$}\\
        \midrule 
            PWT                 & 2.964 & 18 & 0.008 & 0.680\\
            SAM\_Valence       & $-$5.333 & 18 & $<$0.001 & $-$1.224 \\
            SAM\_Arousal        & $-$2.974 & 18 & 0.008 & $-$0.682\\
            SAM\_Dominance      & $-$2.455 & 18 & 0.025 & $-$0.563\\
            UEQS\_Pragmatic	    & $-$5.852 & 18 & $<$0.001 & $-$1.342\\
            UEQS\_Hedonic	    & $-$5.249 & 18 & $<$0.001 & $-$1.204\\
            UEQS\_Overall	    & $-$6.341 & 18 & $<$0.001 & $-$1.455\\
            POQ\_Psychological\_Ownership             & $-$4.716 & 18 & $<$0.001 & $-$1.082\\
            POQ\_Affect              & $-$3.438 & 18 & 0.003 & $-$0.789\\
            SAQ\_Interaction	& $-$2.111 & 18 & 0.049 & $-$0.484\\
            SAQ\_Isolation	    & $-$3.644 & 18 & 0.002 & $-$0.836\\
            SAQ\_Safety	        & 2.348  & 18 & 0.031 & 0.539\\   
        \bottomrule
        \end{tabular}
        \end{threeparttable}
    \label{tab:analysis-t-test-table-setup}
\end{table}

\subsection{Lighting}
\label{sec:results-lighting}
A paired samples t-test was performed to evaluate whether there was a difference between the lighting default and customizable condition. It was found that lighting had a statistically significant effect on the dependent variables user experience (SAM and UEQ-S), psychological ownership (POQ), and social acceptability (SAQ) dimensions. Table \ref{tab:analysis-t-test-lighting} gives an overview of the effects. The lighting was significantly more pleasant in the customizable condition (M = 3.95, SD = 0.62) than in the default condition (M = 2.74, SD = 0.99) (SAM Valence). Furthermore, the average SAM Dominance rating indicated significantly higher user control in the lighting customizable condition (M = 3.53, SD = 0.77) than in the default condition (M = 2.74, SD = 1.20). The average UEQS Pragmatic value was significantly better in the lighting customizable condition (M = 1.96, SD = 0.73) than in the default condition (M = 0.53, SD = 1.04). Furthermore, the average UEQS Hedonic value was significantly better in the lighting customizable condition (M = 0.78, SD = 0.90) than in the default condition (M = $-$1.24, SD = 1.50). Moreover, the average UEQS Overall value was significantly better in the lighting customizable condition (M = 1.37, SD = 0.58) than in the default condition (M = $-$0.36, SD = 1.11). Participants' feelings of ownership toward the lighting (POQ Psychological Ownership) were significantly higher in the lighting customizable condition (M = 5.18, SD = 1.34) than in the default condition (M = 2.09, SD = 0.87). Furthermore, participants' feelings toward the lighting (POQ Affect) were significantly better in the lighting customizable condition (M = 6.16, SD = 0.76) than in the default condition (M = 4.07, SD = 1.33). Participants reported significantly higher acceptability of interactions (SAQ Interaction) in the lighting customizable condition (M = 4.50, SD = 1.44) than in the default condition (M = 3.82, SD = 1.61).

\begin{table}[tb]
    \centering
    \caption{Paired Samples T-Test Statistically Significant Results (\textit{p}$<$0.05) for the Independent Variable Lighting ($N = 19$)}
        \begin{threeparttable}[t]
        \begin{tabular}{lrrrrrr}
        \toprule
        \multicolumn{1}{c}{Dependent Variable} & \multicolumn{1}{c}{$t$} & \multicolumn{1}{c}{$df$}& \multicolumn{1}{c}{$p$}& \multicolumn{1}{c}{$d$}\\
        \midrule 
            SAM\_Valence       & $-$5.115  & 18 & $<$0.001 & $-$1.173\\
            SAM\_Dominance	    & $-$2.535 & 18 & 0.021 & $-$0.582\\
            UEQS\_Pragmatic	    & $-$5.706 & 18 & $<$0.001 & $-$1.309\\
            UEQS\_Hedonic	    & $-$6.537 & 18 & $<$0.001 & $-$1.500\\
            UEQS\_Overall	    & $-$8.200 & 18 & $<$0.001 & $-$1.881\\
            POQ\_Psychological\_Ownership             & $-$8.983 & 18 & $<$0.001 & $-$2.061\\
            POQ\_Affect              & $-$7.390 & 18 & $<$0.001 & $-$1.695\\
            SAQ\_Interaction	& $-$3.256 & 18 & 0.004 & $-$0.747\\   
        \bottomrule
        \end{tabular}
        \end{threeparttable}
    \label{tab:analysis-t-test-lighting}
\end{table}

\section{Discussion and Conclusion}

This investigation explored the effects of customizable augmented table setup and lighting on various aspects of user experience, perceived waiting time, psychological ownership, and social acceptability within a simulated restaurant environment using VR technology. The study aimed to test an initial concept for AR glasses in restaurants, showing potential for academic research and practical applications. The effects of table setup and lighting on UX, perceived waiting time, psychological ownership, and social acceptability were assessed by comparing default and customizable conditions. The results indicated that customizable table setup and lighting significantly enhanced both the pragmatic and hedonic qualities of UX. These enhancements were evident through increased levels of pleasure and, to a lesser extent, arousal and dominance. The ability to customize the simulated restaurant environment to match individual preferences likely contributed to these improvements, fostering a stronger sense of psychological ownership and a deeper connection to the setting \cite{85hassenzahl2007, 33carrozzi2019}. Furthermore, the customizable table setup significantly reduced participants' perceived waiting time. This finding suggests that engaging in AR activities with virtual objects could positively influence customers' experiences in real restaurant settings by altering their perception of time \cite{36hwang2012, 116kim2011}. The study also examined the social acceptability of AR usage in a restaurant setting. Customizable AR interactions were perceived as more socially acceptable compared to passive usage, as active engagement with a customizable table setup could reduce feelings of isolation and promote a more inclusive and engaging experience \cite{87koelle2015}. However, safety concerns were noted, particularly regarding the potential distractions associated with interacting with virtual objects, which could pose risks to physical navigation \cite{128regenbrecht}. Nevertheless, 73.7\% of participants openly stated that they would use AR glasses in an AR Restaurant in a future scenario, and 84.2\% underscored that they would like to experience it with someone else. While the benefits of customization are apparent across multiple dimensions, the study also highlights the complexities of managing safety and acceptability. This study contributes valuable insights into integrating AR technology within the hospitality industry, demonstrating the significant potential of customizable AR enhancements to enrich the restaurant experience by fostering interactivity and personalization. The findings support further exploration of AR applications in restaurant settings and encourage continued academic and practical inquiry to refine and maximize the utility of these technologies \cite{75bilgili2018}. However, the study has limitations that should be addressed in future research. The experiment required participants to imagine other people, and some participants experienced fatigue from the VR headset, which may have influenced their responses. 
In future research, it could be investigated whether there is a threshold for the number of customizable attributes and values, beyond which negative effects on the dependent variables might occur \cite{132park2022}. Additionally, further analysis and comparison of the combined effects of table setup and lighting could be conducted to determine whether these elements together provide additional benefits or present new challenges in enhancing user experience. Also, the shared experience intended in the concept should be investigated, including assessing the variables from this study and extending for social interaction variables. In general, future work could expand this research by exploring themes, different customization options, and scenario variations \cite{56wagener2022}.

\bibliographystyle{IEEEtran}
\bibliography{main}

\end{document}